\documentclass[letter,twocolumn]{jpsj2} %% two-column layout
\newcommand{\Eqn}[1]{&\hspace{-0.5em}#1\hspace{-0.5em}&}
\newcommand{\simg}{\stackrel{>}{_\sim}}
\newcommand{\siml}{\stackrel{<}{_\sim}}

\title{Magnetism and Superconductivity in the Two-Dimensional 16 Band \textit{d}-\textit{p} Model for Iron-Based Superconductors}

\author{Yuki \textsc{Yanagi}\thanks{E-mail: yanagi@phys.sc.niigata-u.ac.jp}, Youichi \textsc{Yamakawa} and Yoshiaki \textsc{\=Ono}}

\inst{Department of Physics, Niigata University, Ikarashi, Niigata
950-2181, Japan \\
}

\abst{The electronic states of the Fe$_2$As$_2$ plane in iron-based
superconductors are investigated on the basis of the two-dimensional
16-band $d$-$p$ model which includes the
Coulomb interaction on a Fe site: the intra- and inter-orbital direct
terms $U$ and $U'$, the Hund's coupling $J$ and the pair-transfer
$J'$. Using the random phase
approximation (RPA), we obtain the magnetic phase diagram including the
stripe and the incommensurate order on the $U'$-$J$ plane. We also solve the superconducting gap equation within the RPA and
find that, for large $J$, the
most favorable pairing symmetry is extended $s$-wave whose
order parameter changes its sign between the hole pockets and the
electron pockets, while it is $d_{xy}$-wave for small $J$.}

\kword{iron-based superconductors, 16-band \textit{d}-\textit{p} model,
pairing symmetry, RPA, Hund's coupling}

\begin{document}
\maketitle
The newly discovered iron-based superconductors\cite{kamihara_1,kamihara}
RFe$Pn$O$_{1-x}$F$_x$ (R=Rare Earth, $Pn$=As, P) with a transition temperature up
to $T_c=55{\rm K}$\cite{ren_3} have attracted much
attention. The F nondoped samples exhibit the stripe-type antiferromagnetic
order with a transition temperature $134\mathrm{K}$ and a magnetic moment
$\sim0.36\mu_{B}$\cite{cruz} at low temperature. With increasing F doping, the system becomes metallic and the
antiferromagnetic order disappears\cite{kamihara}, and then, the
superconductivity emerges for $x\sim0.11$ with $T_c\sim26\mathrm{K}$.
Specific features of the systems are two-dimensionality of the
conducting Fe$_{2}$As$_{2}$ plane and the orbital degrees of freedom in
Fe$^{2+}$ (3$d^6$)\cite{kamihara_1,kamihara}. The pairing symmetry together with the mechanism of the superconductivity is one of the most significant issues. 

The NMR Knight shift measurements
revealed that the superconductivity of the systems is the spin-singlet pairing\cite{zheng,kawabata_1}. Fully
gapped superconducting states have been predicted by various experiments
such as the penetration depth\cite{hashimoto}, the specific heat\cite{mu}, the angle
resolved photoemission spectroscopy (ARPES)\cite{liu,ding,kondo} and the impurity effect on
$T_c$\cite{kawabata_1}. 
 In contrast to the above mentioned experiments, the NMR relaxation rate
 shows the power low behavior $1/T{_1}\propto T^3$ below
 $T_{c}$\cite{nakai}, suggesting the superconducting gaps with line nodes.

Theoretically, the first principle calculations have predicted that the
nondoped system is metallic with two or three concentric hole Fermi surfaces around $\Gamma$
($\mathbf{k}=(0,0)$) point and two elliptical
electron Fermi surfaces around $M$ ($\mathbf{k}=(\pi,\pi)$)
point\cite{singh,haule,xu,boeri}. Mazin \textit{et al.} suggested
that the spin-singlet extended $s$-wave pairing whose order parameter
changes its sign between the hole pockets and the electron pockets is
favored due to the antiferromagnetic spin fluctuations\cite{mazin,seo}.
According to the
weak coupling approaches based on multi-orbital Hubbard models\cite{kuroki,nomura_1,wang_1,yao},
the extended $s$-wave pairing or the $d_{xy}$-wave
pairing is expected to emerge. The details of the band structure and the Fermi
surface are crucial for determining the pairing symmetry. Therefore,
theoretical studies based on a more realistic model which includes both the Fe $3d$ orbitals
and the As $4p$ orbitals, so called $d$-$p$ model, are highly desired. 

In the previous paper\cite{yamakawa}, we have investigated the pairing symmetry of the
two-dimensional 16-band $d$-$p$ model by using the random phase
approximation (RPA). It has been found that, for a larger value of $J/U'$,
the most favorable paring symmetry is extended $s$-wave, while, for a
smaller value of $J/U'$, it is $d_{xy}$-wave. However, the detailed
electronic states in the whole parameter region of $U'$ and $J$ have not
been discussed there. The purpose of this paper is to obtain the
detailed phase diagram including the magnetism and the superconductivity
in the $U'$-$J$ plane.

 First of all, we perform the density functional calculation for LaFeAsO with the
 generalized gradient approximation of Perdew, Burke and Ernzerhof\cite{perdew} by using the WIEN2k
package\cite{blaha}, where the lattice parameters ($a=4.03268$\AA, $c=8.74111$\AA) and
the internal coordinates ($z_{La}=0.14134$, $z_{As}=0.65166$) are
experimentally determined\cite{nomura}. Considering that there are two distinct Fe and As
sites in the crystallographic unit cell, we
then derive the two-dimensional 16-band $d$-$p$ model\cite{yamakawa,cvetkovic}, where $3d$ orbitals ($d_{3z^2-r^2}$, $d_{x^2-y^2}$, $d_{xy}$, $d_{yz}$, $d_{zx}$) of two Fe
atoms (Fe$^1$=$A$, Fe$^2$=$B$) and $4p$ orbitals ($p_{x}$, $p_{y}$, $p_{z}$) of two As atoms are
explicitly included. We note that
$x, y$ axes are rotated by 45 degrees from the direction along Fe-Fe
bonds. The noninteracting part of the $d$-$p$ model is given by the following tight-binding Hamiltonian,
\begin{eqnarray}
%H\Eqn{=}H_0+H_{\mathrm{int}},\\
H_0\Eqn{=}\sum_{i,\ell,\sigma}\hspace{-1mm}
\varepsilon^d_{\ell}d^{\dag}_{i\ell\sigma}d_{i\ell\sigma}
+\hspace{-1mm}\sum_{i,m,\sigma}\hspace{-1mm}\varepsilon^p_{m}p^{\dag}_{im\sigma}p_{im\sigma} \nonumber \\ 
\Eqn{+}\sum_{i,j,\ell,\ell',\sigma}\hspace{-1mm}t^{dd}_{i,j,\ell,\ell'}d^{\dag}_{i\ell\sigma}d_{j\ell'\sigma}
 +\hspace{-1mm} \sum_{i,j,m,m',\sigma}\hspace{-1mm}t^{pp}_{i,j,m,m'}p^{\dag}_{im\sigma}p_{jm'\sigma} \nonumber\\
\Eqn{+}\sum_{i,j,\ell,m,\sigma}\hspace{-1mm}t^{dp}_{i,j,\ell,m}d^{\dag}_{i\ell\sigma}p_{jm\sigma}+h.c. \label{d-p}, 
\end{eqnarray}
where $d_{i\ell\sigma}$ is the annihilation operator for Fe-$3d$ electrons with spin
$\sigma$ in the orbital $\ell$ at the site $i$ and $p_{im\sigma}$ is the annihilation
operator for As-$4p$ electrons with spin
$\sigma$ in the orbital $m$ at the site $i$. In eq. (2), the
transfer integrals $t^{dd}_{i,j,\ell,\ell'}$, $t^{pp}_{i,j,m,m'}$,
$t^{dp}_{i,j,\ell,m}$ and the atomic energies $\varepsilon^d_{\ell}$,
$\varepsilon^p_{m}$ are determined so as to fit both the energy and the weights of orbitals for each band obtained from the
tight-binding approximation to
those from the density functional calculation\cite{yamakawa_2}. The doping $x$
corresponds to the number of electrons per unit cell $n=24+2x$ in the present
model. 
%%%%%%%%%%%%%%%%%%%%%%%%%%%%% BAND STRUCTURE AND FERMI SURFACE %%%%%%%%%%%%%%%%%%%%%%%%%%%%%%%%%%%%
\begin{figure}[b]
\begin{center}
\begin{minipage}{40mm}
\begin{center}
\includegraphics[width=4cm]{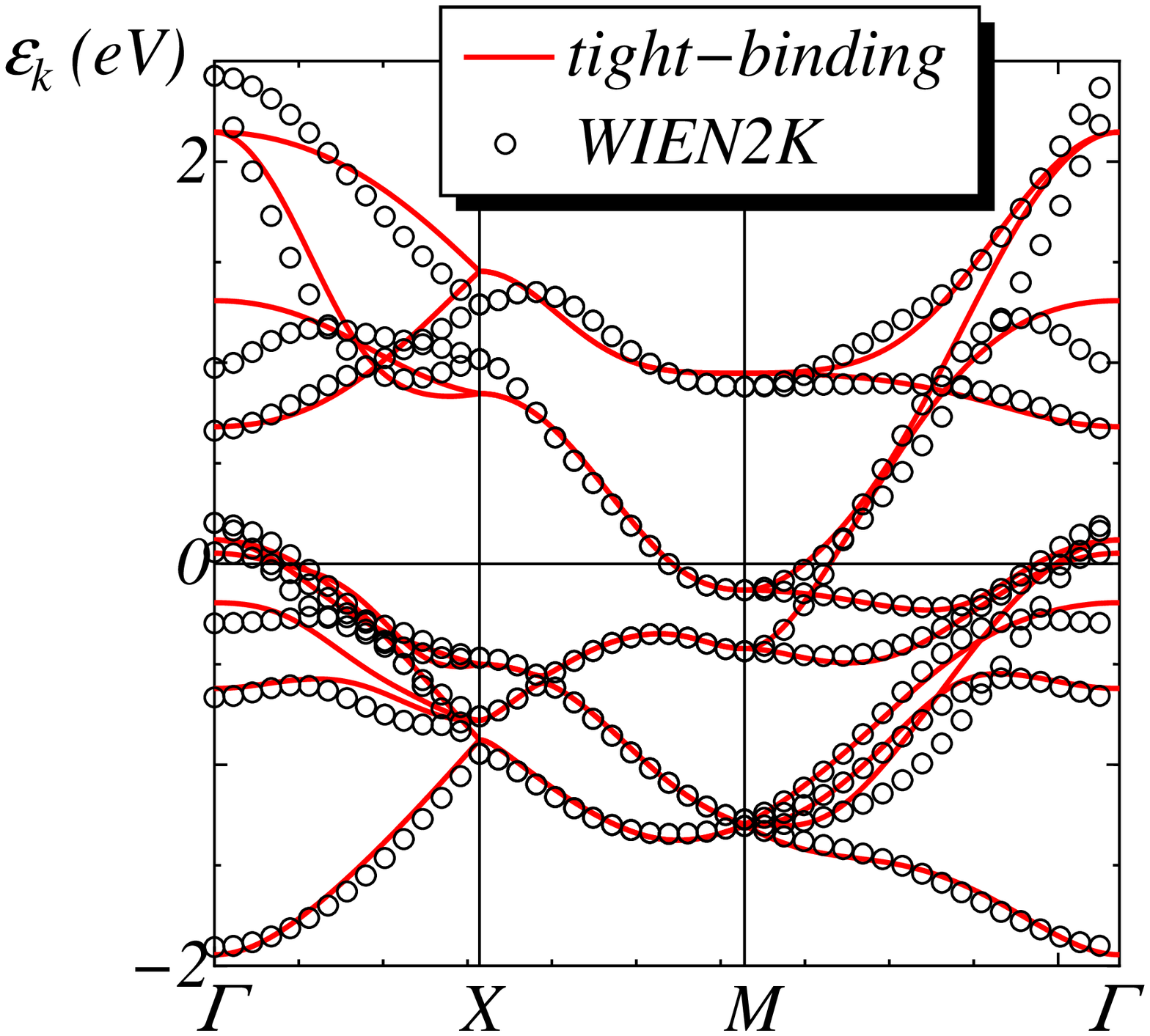}
\end{center}
\end{minipage}
\hspace{5mm}
\begin{minipage}{35mm}
\begin{center}
\includegraphics[width=3.5cm]{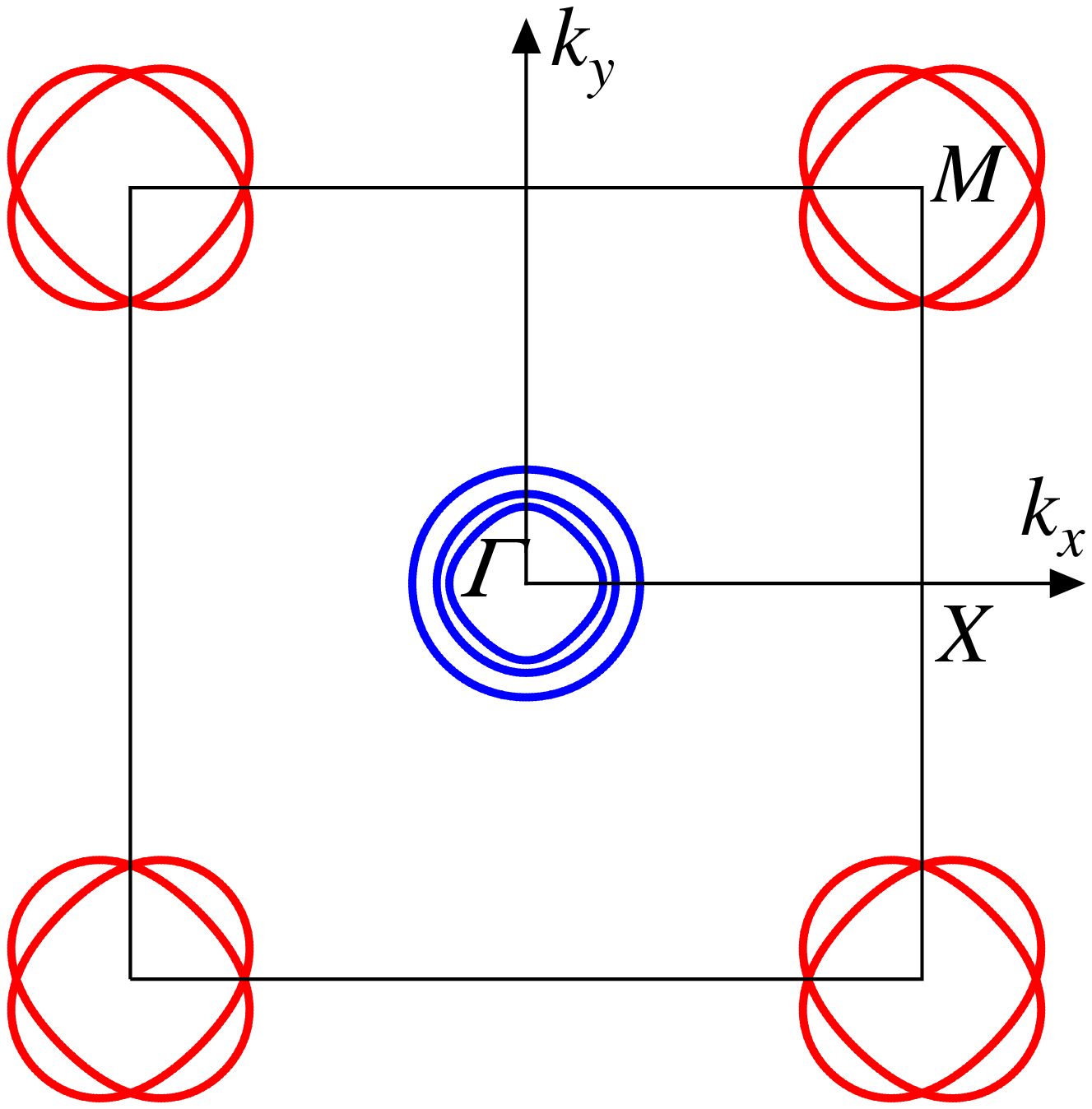}
\end{center}
\end{minipage}
\end{center}
\vspace{-3.5mm}
\caption{(Color online) (Left panel) The band structure obtained from eq. (\ref{d-p})
 (solid line)
 and that obtained from the density functional calculation (open
 circle). 
(Right panel) Fermi surface obtained from the $d$-$p$ model eq. (\ref{d-p}) for $x=0$ \label{FS}}
\end{figure}
%%%%%%%%%%%%%%%%%%%%%%%%%%%%%%%%%%%%%%%%%%%%%%%%%%%%%%%%%%%%%%%%%%%%%%%%%%%%%%%%%%%%%%%%%%%%%%%%%%%%
%%%%%%%%%%%%%%%%%%%%%%%%%%%%%% NONINTERACTING SUSCEPTIBILITY
%%%%%%%%%%%%%%%%%%%%%%%%%%%%%%%%%%%%%%%
%\vspace{-10mm}
\begin{figure}
\begin{center}
\vspace{-5mm}
\begin{minipage}{35.0mm}
\begin{center}
\includegraphics[width=3.5cm]{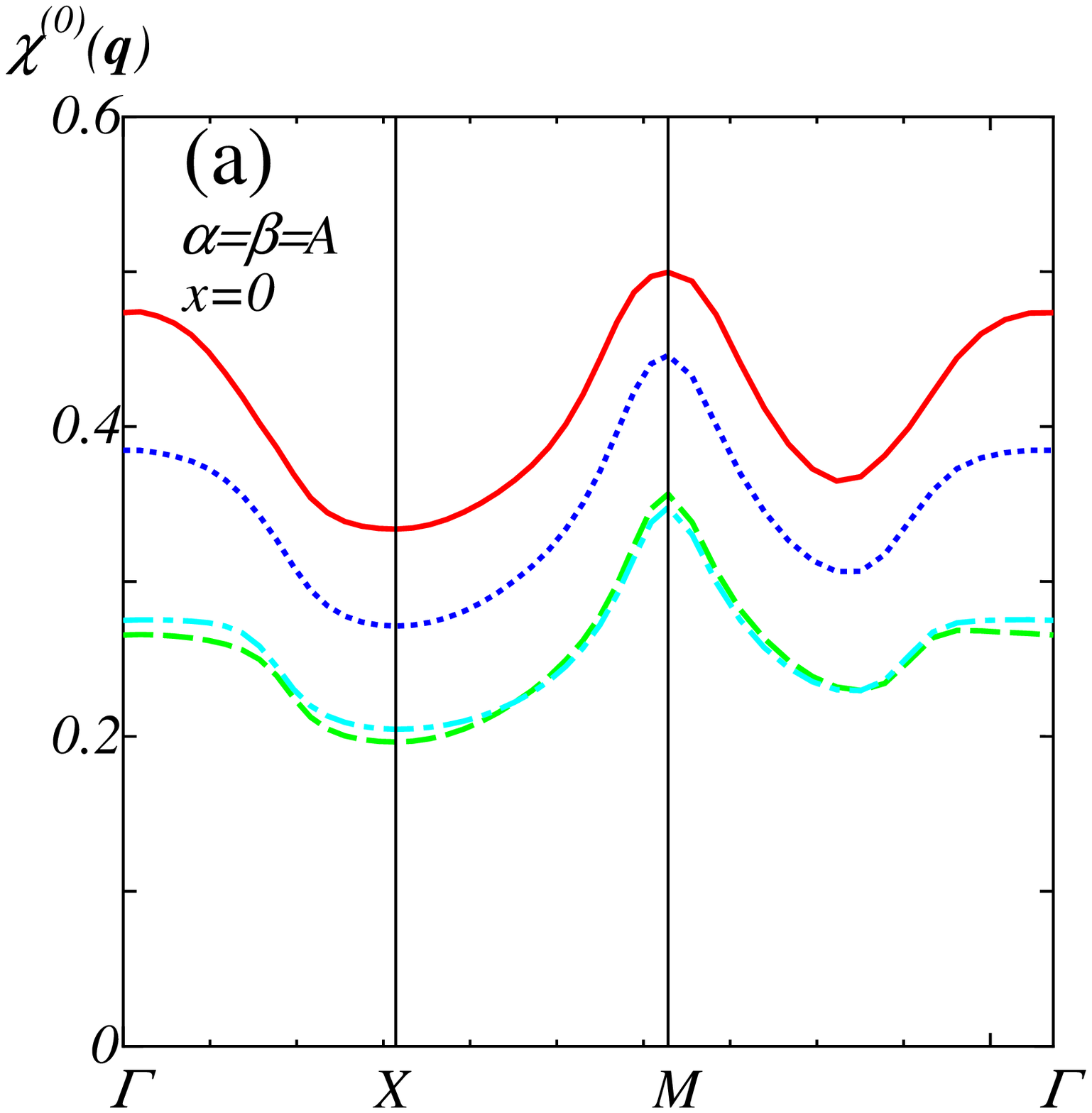}
\end{center}
\end{minipage}
\begin{minipage}{35.0mm}
\begin{center}
\includegraphics[width=3.5cm]{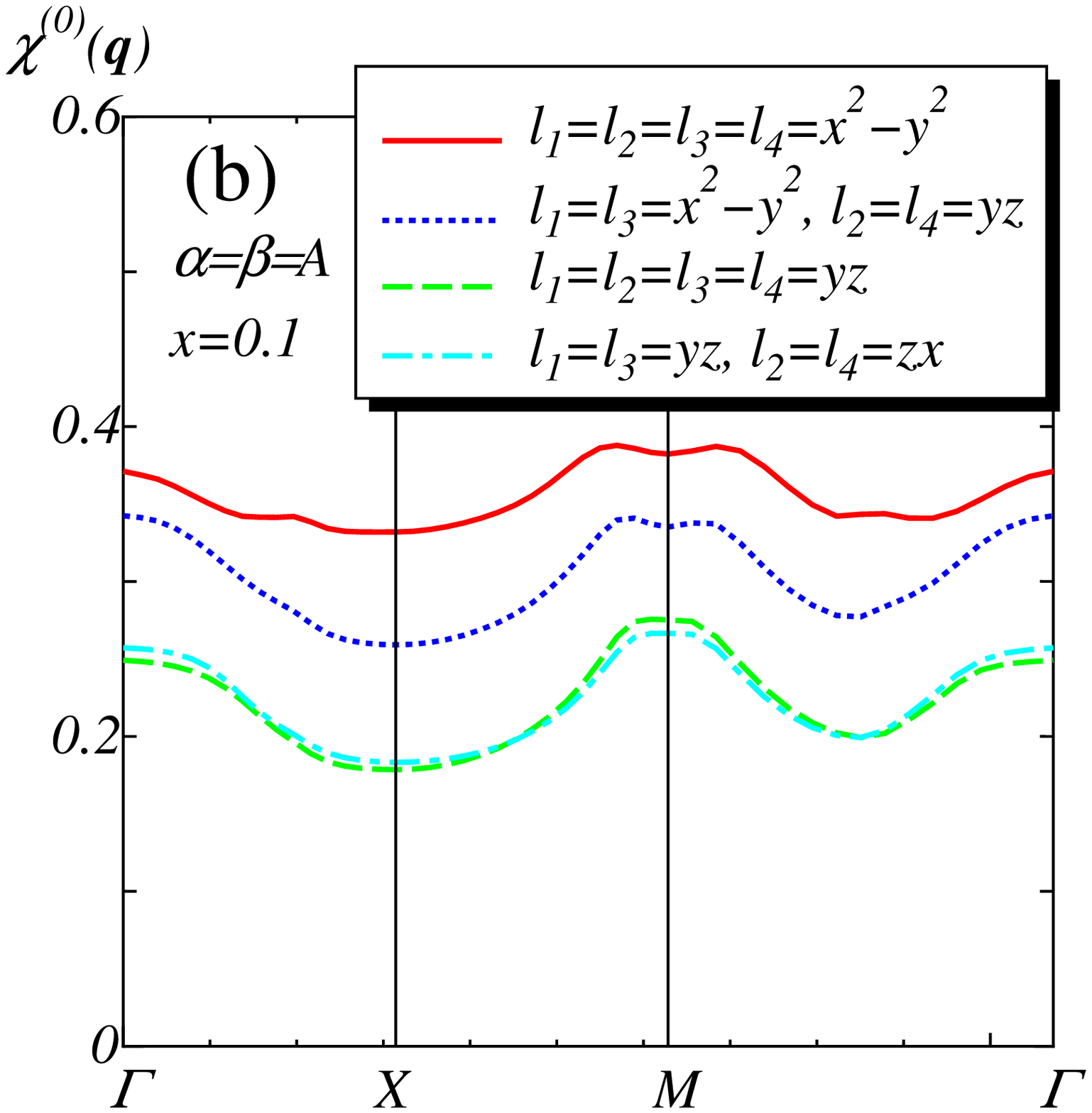}
\end{center}
\end{minipage}
\caption{(Color online) Several components of the noninteracting susceptibility $\chi^{(0)~\alpha\beta}_{\ell_1\ell_2,\ell_3\ell_4}$ for
 $x=0$ (a) and those for $x=0.1$ (b). \label{chi0}}
\end{center}
\end{figure}
%%%%%%%%%%%%%%%%%%%%%%%%%%%%%%%%%%%%%%%%%%%%%%%%%%%%%%%%%%%%%%%%%%%%%%%%%%%%%%%%%%%%%%%%%%%%%%%%%%%%%

We show the band structure obtained from the $d$-$p$ tight-binding
Hamiltonian eq. (\ref{d-p}) together with that
obtained from
the density functional calculation in the left panel of
Fig. \ref{FS}. It is found that the former reproduces the latter very
well. We note that the weights of orbitals also agree very well with each other
(not shown). The 10 bands near the Fermi level are mainly constructed by the Fe
$3d$ orbitals and the 6 bands below the $3d$ 10 bands are mainly
constructed by the As $4p$ orbitals (not shown in Fig. \ref{FS}). The Fermi surface
for the $d$-$p$ tight-binding model is shown in the right panel of Fig. \ref{FS}, where
we can see nearly circular hole pockets around the $\Gamma$
point  and elliptical electron pockets around the
$M$ point. These results are consistent with
the first principle calculations\cite{singh,haule,xu,boeri}. 

Now we consider the effect of the Coulomb interaction on Fe site: the
intra-orbital (inter-orbital) direct terms $U$ ($U'$), 
the Hund's rule coupling $J$ and the pair-transfer $J'$. 
Within the RPA\cite{takimoto}, the spin susceptibility $\hat{\chi^s}(\mathbf{q})$ and the charge-orbital
susceptibility $\hat{\chi^c}(\mathbf{q})$ are given in the $50\times50$
matrix representation as follows, 
\begin{eqnarray}
\hat{\chi^s}(\mathbf{q})\Eqn{=}(\hat{1}-\hat{\chi}^{(0)}(\mathbf{q})\hat{S})^{-1}\hat{\chi}^{(0)}(\mathbf{q}) \label{eq_chis},\\
\hat{\chi^c}(\mathbf{q})\Eqn{=}(\hat{1}+\hat{\chi}^{(0)}(\mathbf{q})\hat{C})^{-1}\hat{\chi}^{(0)}(\mathbf{q}) \label{eq_chic}
\end{eqnarray}
with the noninteracting susceptibility
\begin{eqnarray}
\Eqn{}\chi^{(0)~\alpha,\beta}_{\ell_1\ell_2,\ell_3\ell_4}(\mathbf{q})=
-\frac{1}{N}\sum_{\mathbf{k}}\sum_{\mu,\nu}
\frac{f(\varepsilon_{\mathbf{k}+\mathbf{q},\mu})
-f(\varepsilon_{\mathbf{k},\nu})}
{\varepsilon_{\mathbf{k}+\mathbf{q},\mu}
-\varepsilon_{\mathbf{k},\nu}} \nonumber\\
\Eqn{\times}u^{\alpha}_{\ell_1,\nu}(\mathbf{k})^*
{u^{\alpha}_{\ell_2,\mu}(\mathbf{k}+\mathbf{q})}
u^{\beta}_{\ell_3,\nu}(\mathbf{k})
{u^{\beta}_{\ell_4,\mu}(\mathbf{k}+\mathbf{q})}^* \label{eq_chi0},
\end{eqnarray}
where $\mu$, $\nu$ (=1-16) are band indexes, $\alpha$, $\beta$ ($=$$A,B$) represent two
Fe sites, $\ell$ represents Fe 3$d$ orbitals,
$u^{\alpha}_{\ell,\mu}(\mathbf{k})$ is the eigenvector which
diagonalizes $H_0$ eq. (\ref{d-p}), $\varepsilon_{\mathbf{k},\mu}$ is
the corresponding eigenenergy of band
$\mu$ with wave vector $\mathbf{k}$ and $f(\varepsilon)$ is the Fermi
distribution function. In eqs. (2) and (3), the interaction matrix $\hat{S}$ ($\hat{C}$) is given by
\begin{equation}
\hat{S}~(\hat{C})= \left\{
\begin{array}{@{\,} l @{\,} c}
U~(U) & (\alpha=\beta,~\ell_1=\ell_2=\ell_3=\ell_4)\\
U'~(-U'+2J) & (\alpha=\beta,~\ell_1=\ell_3\ne\ell_2=\ell_4)\\
J~(2U'-J) & (\alpha=\beta,~\ell_1=\ell_2\ne\ell_3=\ell_4)\\
J'~(J')& (\alpha=\beta,~\ell_1=\ell_4\ne\ell_2=\ell_3)\\
0 & (\mathrm{otherwise})
\end{array} \right. .\nonumber
\end{equation} 

In the weak coupling regime, the superconducting gap equation is given by 
\begin{eqnarray}
\Eqn{}\lambda\Delta^{\alpha\beta}_{\ell\ell'}({\bf k})=\frac{1}{N}\sum_{{\bf k}'}\sum_{\ell_1\ell_2\ell_3\ell_4}\sum_{\alpha',\beta'}\sum_{\mu,\nu}\ \ \ \ \ \ \ \ \ \ \ \ \ \ \nonumber\\
\Eqn{\times}\frac{f(\varepsilon_{-{\bf k}',\mu})+f(\varepsilon_{{\bf k}',\nu})-1}{\varepsilon_{-{\bf k}',\mu}+\varepsilon_{{\bf k}',\nu}}V^{\alpha,\beta}_{\ell\ell_1,\ell_2\ell'}({\bf k}-{\bf
 k}'){\Delta}^{\alpha'\beta'}_{\ell_3\ell_4}({\bf k}')\nonumber\\
\Eqn{\times}u^{\alpha'}_{\ell_3,\mu}(-{\bf k}'){u^{\alpha}_{\ell_1,\mu}(-{\bf k}')}^*u^{\beta'}_{\ell_4,\nu}({\bf k}'){u^{\beta}_{\ell_2,\nu}({\bf k}')}^* \label{eq_gap},
\end{eqnarray}
where $\Delta^{\alpha\beta}_{\ell\ell'}({\bf k})$ is the gap function
and $V^{\alpha,\beta}_{\ell_1\ell_2,\ell_3\ell_4}(\mathbf{q})$ is the
effective pairing interaction.
Within the RPA\cite{takimoto}, $V^{\alpha,\beta}_{\ell_1\ell_2,\ell_3\ell_4}(\mathbf{q})$ 
is given in the $50\times50$
matrix,
\begin{equation}
\hat{V}(\mathbf{q})=\eta(\hat{S}\hat{\chi}^s(\mathbf{q})\hat{S}+\frac{1}{2}\hat{S})-\frac{1}{2}(\hat{C}\hat{\chi}^c(\mathbf{q})\hat{C}-\frac{1}{2}\hat{C})\label{eq_veff_s},
\end{equation}
with $\eta=\frac{3}{2}$ for the spin-singlet state and
$\eta=-\frac{1}{2}$  for the spin-triplet state.
The gap equation eq. (5) is solved to
obtain the gap function $\Delta^{\alpha\beta}_{\ell\ell'}(\mathbf{k})$
with the eigenvalue $\lambda$. At $T=T_c$, the largest eigenvalue $\lambda$
becomes unity.
%%%%%%%%%%%%%%%%%%%%%%%%%%%%%% EFFECTIVE PAIRING INTERACTION %%%%%%%%%%%%%%%%%%%%%%%%%
\begin{figure}
\begin{center}
\begin{minipage}{35.0mm}
\begin{center}
\includegraphics[width=3.5cm]{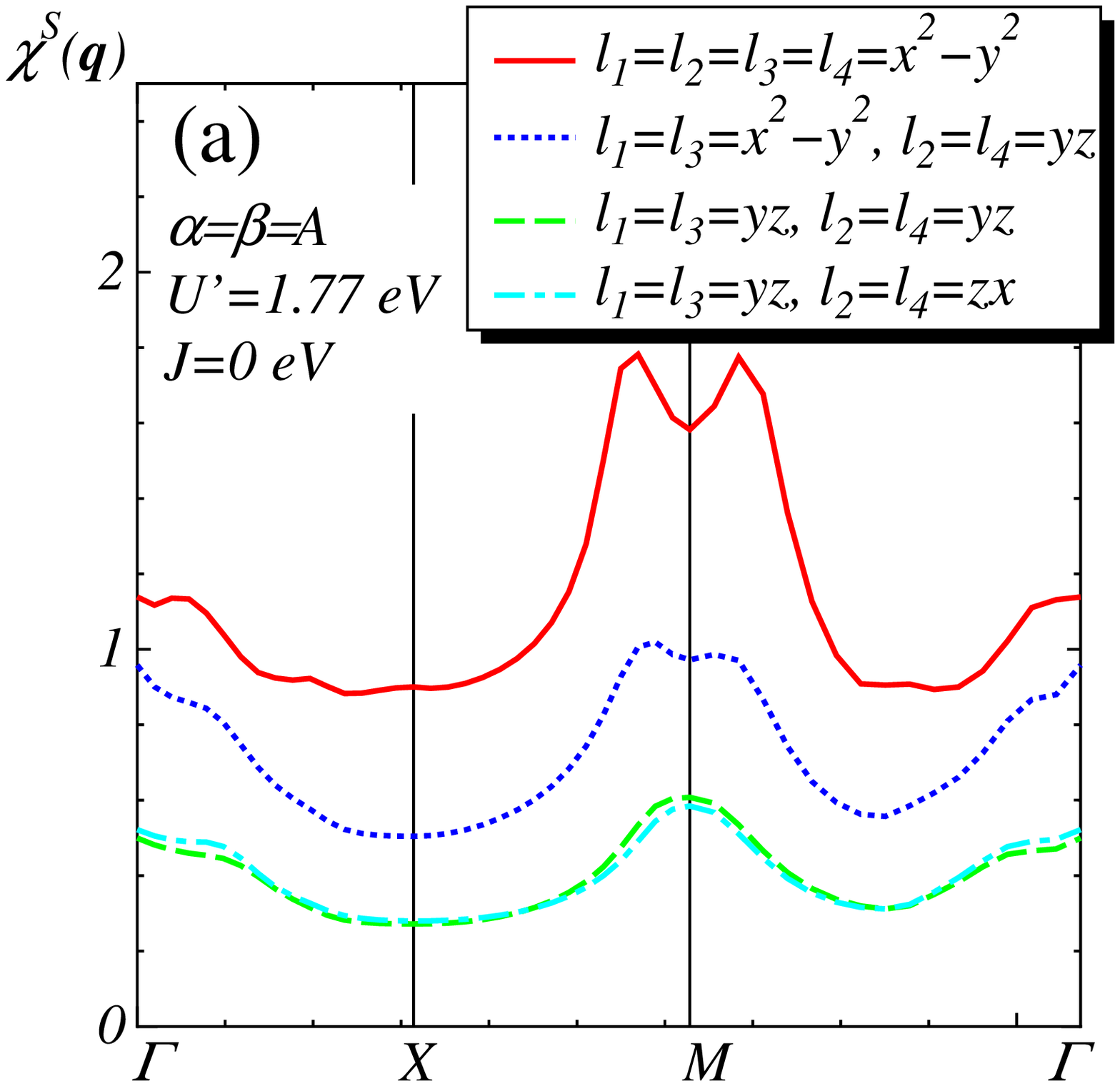}
\end{center}
\end{minipage}
\begin{minipage}{35.0mm}
\begin{center}
\includegraphics[width=3.5cm]{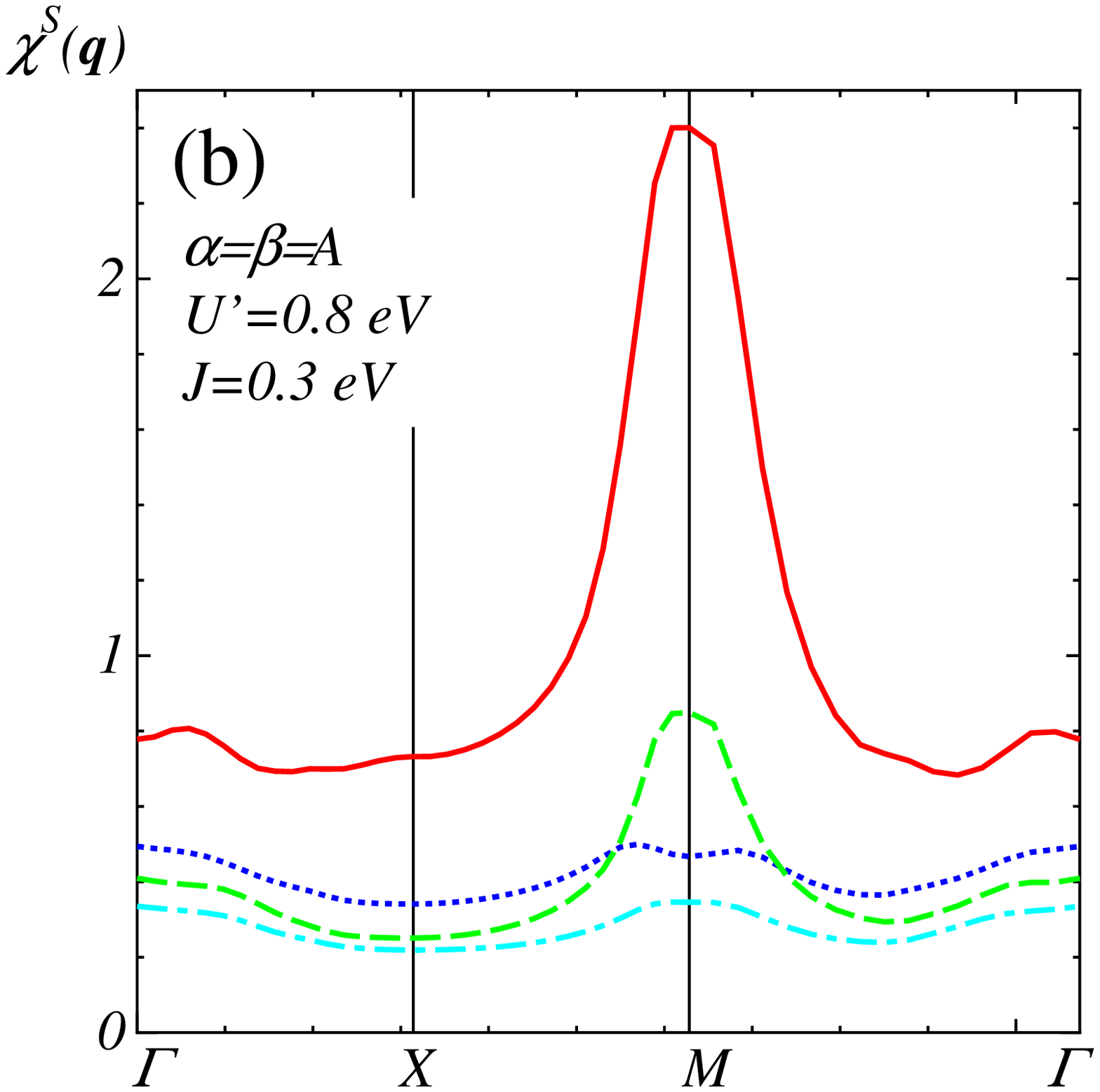}
\end{center}
\end{minipage}
\begin{minipage}{35.0mm}
\begin{center}
\includegraphics[width=3.5cm]{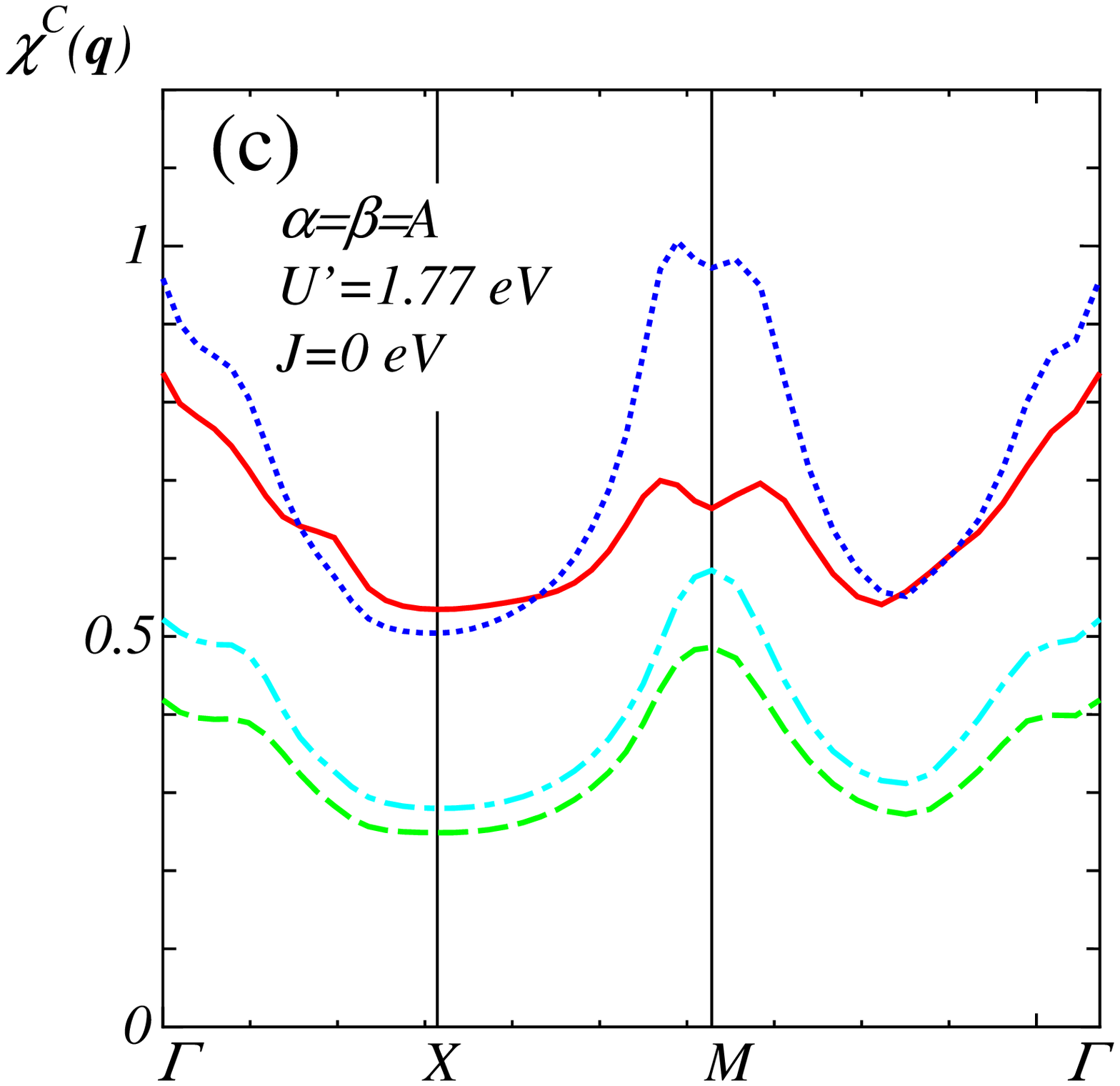}
\end{center}
\end{minipage}
\begin{minipage}{35.0mm}
\begin{center}
\includegraphics[width=3.5cm]{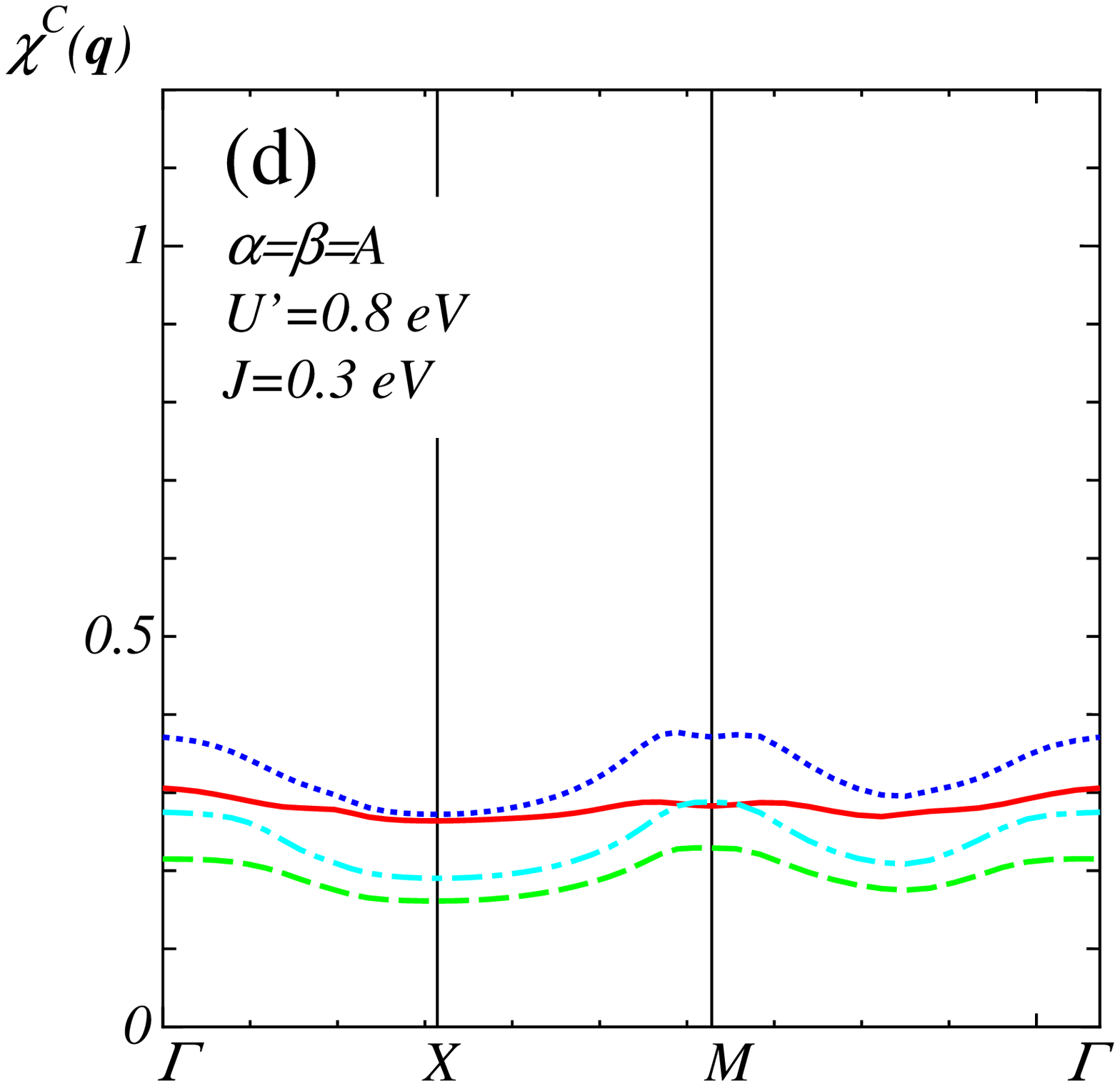}
\end{center}
\end{minipage}
\begin{minipage}{35.0mm}
\begin{center}
\includegraphics[width=3.5cm]{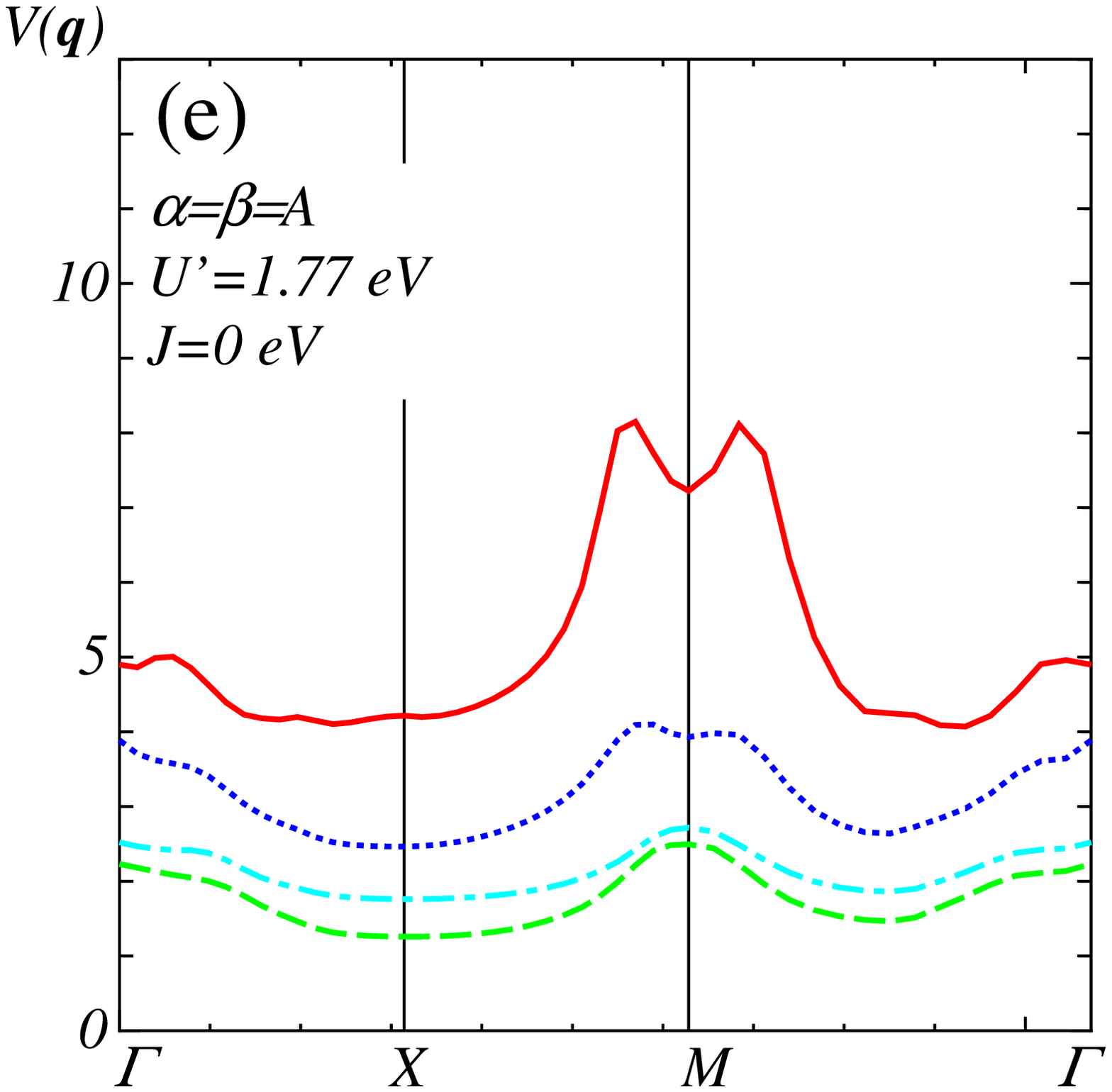}
\end{center}
\end{minipage}
\begin{minipage}{35.0mm}
\begin{center}
\includegraphics[width=3.5cm]{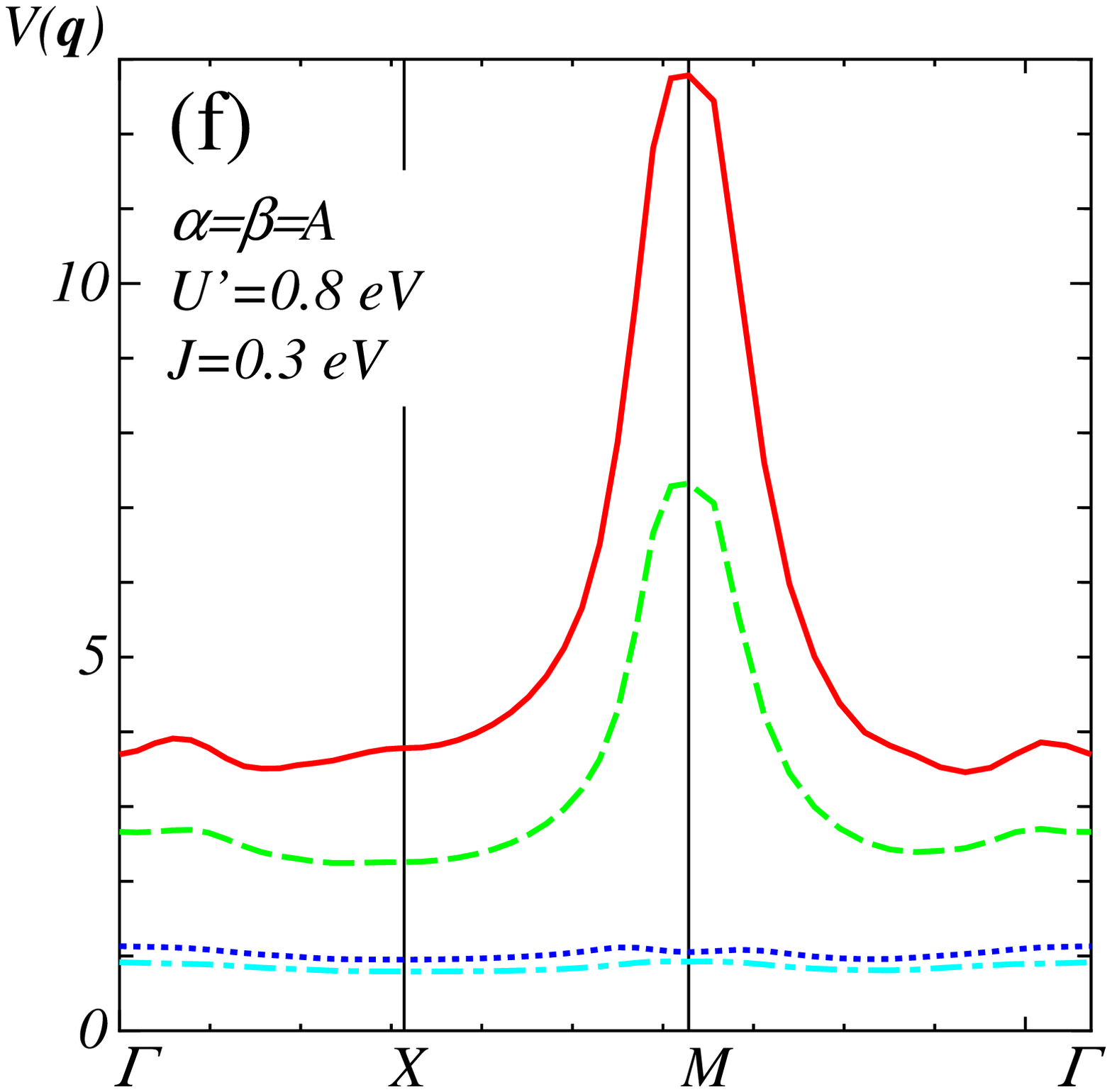}
\end{center}
\end{minipage}
\caption{(Color online) Several components of the spin susceptibility
 $\chi^{s~\alpha,\beta}_{\ell_1\ell_2,\ell_3\ell_4}$ (a) and (b), the
 charge-orbital susceptibility
 $\chi^{c~\alpha,\beta}_{\ell_1\ell_2,\ell_3\ell_4}$ (c) and (d) and
 the effective pairing interaction
 $V^{\alpha,\beta}_{\ell_1\ell_2,\ell_3\ell_4}$ (e) and (f), for
 $U'=1.77\mathrm{eV}$ and $J=0\mathrm{eV}$ (a) (c) (e), and for
 $U'=0.8\mathrm{eV}$ and $J=0.3\mathrm{eV}$ (b) (d) (f), respectively. \label{veff}}
\end{center}
\end{figure}
%\begin{center}
%\includegraphics[width=8.0cm]{veff_u2_177_uj1_0.eps}
%\includegraphics[width=8.0cm]{veff_u2_138_uj1_015.eps}
%\begin{minipage}[30mm]
%\begin{center}
%\includegraphics[width=3.0cm]{chis_u2_080_uj1_030.eps}
%\end{center}
%\end{minipage}
%\end{center}
%\end{figure}
%%%%%%%%%%%%%%%%%%%%%%%%%%%%%%%%%%%%%%%%%%%%%%%%%%%%%%%%%%%%%%%%%%%%%%%%%%%%%%%%%%%%%%
%%%%%%%%%%%%%%%%%%%%%%% EIGENVALUE AND GAP FUNCTION %%%%%%%%%%%%%%%%%%%%%%%%%%%%%%%%%
\begin{figure}[t]
%\begin{minipage}{62mm}
\begin{center}
\includegraphics[width=6.2cm]{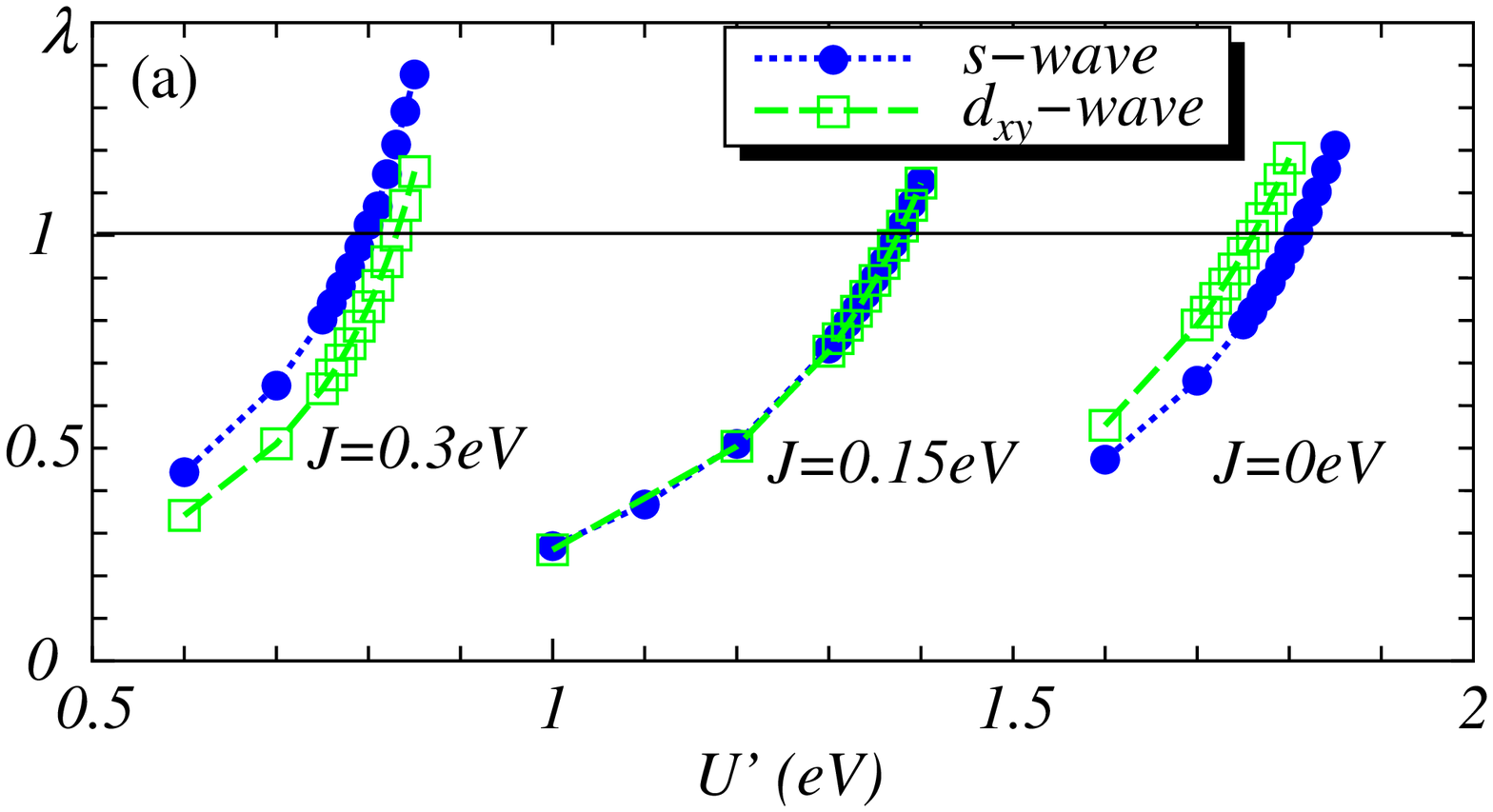}
%\hspace{50mm}
\end{center}
%\end{minipage}
%\vspace{-1cm}
\begin{center}
\begin{minipage}{40mm}
\small{(b)}
\begin{center}
\vspace{-1mm}
\includegraphics[width=4.0cm]{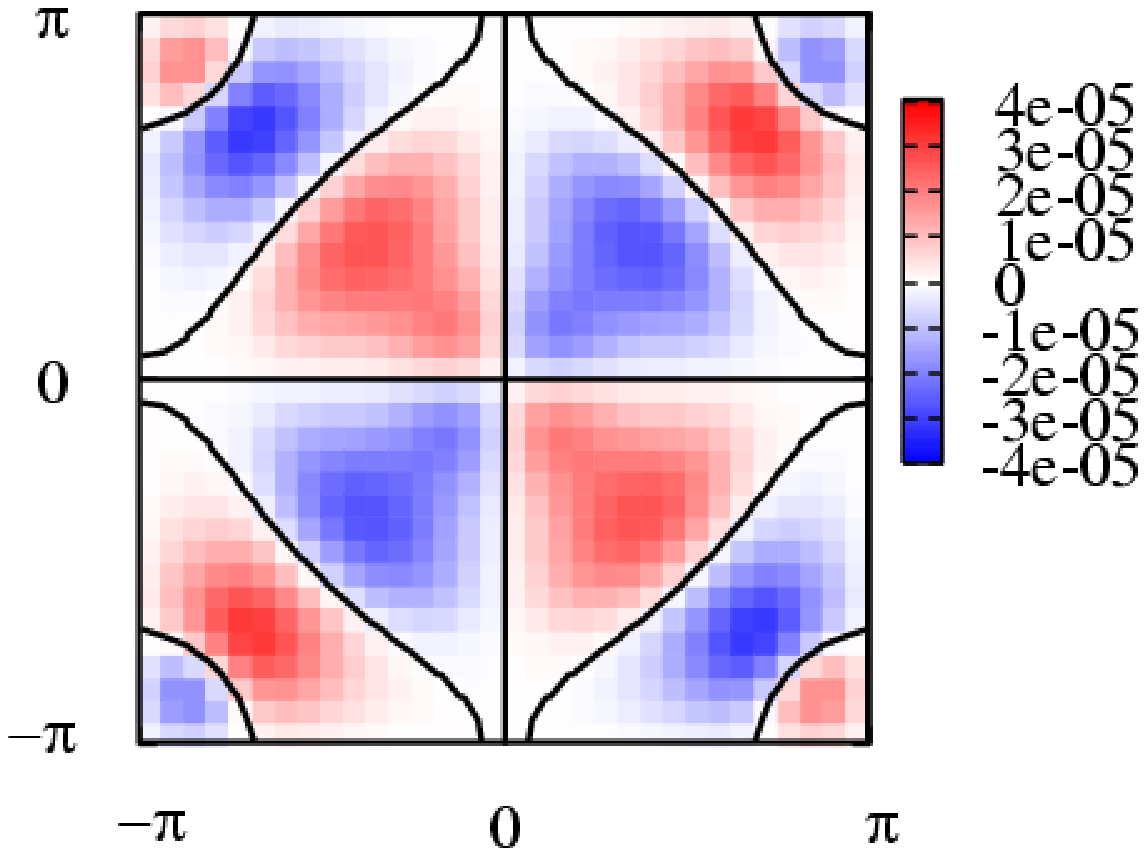}
\end{center}
\end{minipage}
\begin{minipage}{42mm}
\small{(c)}
\begin{center}
\vspace{-1mm}
%\hspace{-2.5cm}
\includegraphics[width=4.2cm]{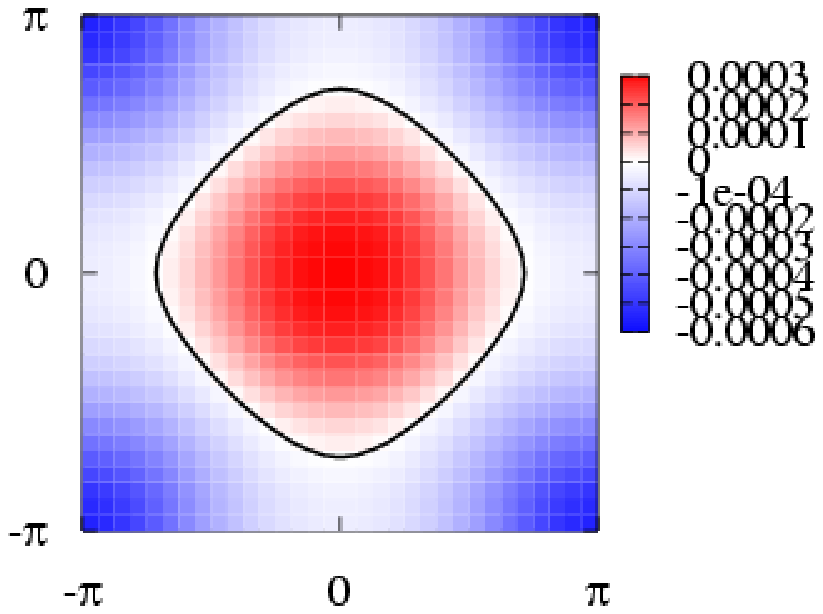}
\end{center}
\end{minipage}
\end{center}
%\hspace{3.5cm}
%\end{center}
\caption{(Color online) (a) $U'$-dependence of the eigenvalue $\lambda$ for $x=0.1$ and 
 $T=0.02$eV. The closed circles and opened squares represent $\lambda$ of the $s$-wave
 and $d_{xy}$-wave, respectively. The gap function
 $\Delta^{AA}_{x^2-y^2,x^2-y^2}(\mathbf{k})$ for $d_{xy}$-wave symmetry
 at $U'=1.77\mathrm{eV}, J=0\mathrm{eV}$ and $x=0.1$ (b), and that for
 $s$-wave symmetry at $U'=0.8\mathrm{eV}, J=0.3\mathrm{eV}$ and $x=0.1$ (c). The solid lines represent the nodes of the gap function.  \label{eigenval}}
\end{figure}
%%%%%%%%%%%%%%%%%%%%%%%%%%%%%%%%%%%%%%%%%%%%%%%%%%%%%%%%%%%%%%%%%%%%%%%%%%%%%%%%%%%%%%

In the present paper, we mainly focus on the case with $x=0.1$, where
the superconductivity is observed in the compounds\cite{kamihara}.
For simplicity, we set $T=0.02\mathrm{eV}$ and $U=U'+2J$, $J=J'$. We use $32\times32$
$\mathbf{k}$ points in the numerical calculations for eqs. (2)-(6), and also use the first Fourier
transformation (FFT) to
solve the gap equation eq. (\ref{eq_gap}).

Fig. \ref{chi0} shows the several components of the noninteracting susceptibility given in eq. (\ref{eq_chi0}) for
$x=0.1$ together with those for $x=0$ for comparison.  
We see the peaks
centered at the $\Gamma$ point and those at the $M$ point for $x=0$,
where the former peaks are due to the nesting between the hole (electron) pockets and
the latter peaks are due to the nesting between the hole pockets
and the electron pockets. With the electron doping, the hole pockets
around the $\Gamma$ point shrink and the smallest one disappears for $x=0.1$, while
the electron pockets become larger (see
Fig. 1 of ref. 23). As the
result, the nesting effect becomes weak, and then, the peak at the $M$
point is suppressed and shifts around the $M$ point for $x=0.1$,
resulting in the incommensurate SDW as mentioned later. 

The several components of the spin susceptibility 
$\chi^{s~\alpha,\beta}_{\ell_1\ell_2,\ell_3\ell_4}(\mathbf{q})$ given in
eq. (\ref{eq_chis}) are
plotted in Figs. \ref{veff} (a) and (b), where the
parameters $U'$ and $J$ are set to the condition with $\lambda=1$
(see Fig. \ref{eigenval} (a)). The spin susceptibility is enhanced due
to the effect of the
Coulomb interaction, especially for the diagonal component of
$d_{x^2-y^2}$. For $J=0\mathrm{eV}$, the incommensurate peaks around $M$ point
are observed as reflecting the structure of the bare susceptibility
shown in Fig. \ref{chi0} (b). On the other hand, for $J=0.3\mathrm{eV}$, the commensurate peaks centered at the $M$
point are observed; which is due to an effect of the Hund's coupling $J$. %The ratio of
%inter-orbital Coulomb interaction to intra-orbital one $U'/U$ decrease with increasing
%$J$. 
%Therefore, for $J=0.3\mathrm{eV}$, the inter-orbital
%components are smaller than those for $J=0\mathrm{eV}$ and the
%intra-orbital components are enhanced, where the peaks around $M$ point becomes more dominant and
%commensurate. 

The several components of the charge-orbital susceptibility
$\chi^{c~\alpha,\beta}_{\ell_1\ell_2,\ell_3\ell_4}(\mathbf{q})$ given in
eq. (\ref{eq_chic}) are
plotted in Figs. (\ref{veff}) (c) and (d). In contrast to the case with
the spin susceptibility, the off-diagonal component of
$d_{x^2-y^2}-d_{yz}$ which corresponds to the orbital susceptibility 
becomes most dominant due to the effect of the inter-orbital Coulomb
interaction $U'$. For
$J=0\mathrm{eV}$,  the orbital susceptibility is enhanced and shows peaks
around the $M$ point together with those at the $\Gamma$ point. On the
other hand, for $J=0.3\mathrm{eV}$, the enhancement of  the orbital
susceptibility is very small, where $U'$
 is smaller than the intra-orbital interaction $U=U'+2J$ which suppresses
the orbital susceptibility.

The several components of the effective pairing interaction
$V^{\alpha,\beta}_{\ell_1\ell_2,\ell_3\ell_4}(\mathbf{q})$ for the spin-singlet state given in eq. (\ref{eq_veff_s}) are
plotted in Figs. \ref{veff} (e) and (f). Since the largest eigenvalue $\lambda$ is always
spin-singlet state in the present study, we show the effective pairing
interaction only for the spin-singlet state. Their structures are
similar to those of the spin susceptibility in the both cases for
$J=0\mathrm{eV}$ and $J=0.3\mathrm{eV}$. This is because the
contributions of the spin
fluctuations to the effective pairing interaction is three times larger
than those of the orbital fluctuations according to
eq. (\ref{eq_veff_s}). 

Substituting $V^{\alpha\beta}_{\ell_1\ell_2,\ell_3\ell_4}(\mathbf{q})$ into the gap
equation eq. (\ref{eq_gap}), we obtain the gap function
$\Delta^{\alpha\beta}_{\ell\ell'}(\mathbf{k})$ with the eigenvalue
$\lambda$.  In Fig. \ref{eigenval} (a), the two largest eigenvalues
$\lambda$, which are for the extended $s$-wave 
and the $d_{xy}$-wave pairing symmetries, are plotted as functions of
$U'$ for several values of $J$. We confirmed that eigenvalues for the other paring symmetries
are much smaller. With
increasing $U'$, $\lambda$ monotonically increases and
finally becomes unity at a critical value $U'_c$ above which the
superconducting state is realized.
 For $J=0\mathrm{eV}$, 
the largest eigenvalue $\lambda$ is for the $d_{xy}$-wave symmetry, where the
gap function has line nodes on the Fermi surfaces as shown in
Fig. (\ref{eigenval}) (b).
On the other hand, for $J=0.3\mathrm{eV}$, 
the largest eigenvalue $\lambda$ is for the extended $s$-wave symmetry, where the
gap function changes its sign between the hole pockets and the electron
pockets without nodes on the Fermi surfaces as shown in
Fig. (\ref{eigenval}) (c)\cite{mazin,kuroki,nomura_1}. 
For $J=0.15\mathrm{eV}$, these two eigenvalues are almost degenerate. 
The extended $s$-wave pairing is mediated by the pairing interaction
with the sharp peak at
$\mathbf{Q}=(\pi,\pi)$ (see Fig. \ref{veff} (f)), while the
$d_{xy}$-wave pairing is mediated by the pairing interaction with the
incommensurate peaks around $\mathbf{Q}=(\pi,\pi)$ together with the
peak at $\mathbf{Q}=(0,0)$ (see Fig. \ref{veff} (e)).
%The gap function for $x^2-y^2$ component $\Delta^{AA}_{x^2-y^2,x^2-y^2}(\mathbf{k})$
%is shown in
%Fig. \ref{eigenval} (b) and (c). This component is most dominant for $s$-wave
%symmetry and is second most dominant for $d_{xy}$-wave symmetry. The most dominant
%component for $d_{xy}$-wave symmetry is
%$\Delta^{AB}_{x^2-y^2,x^2-y^2}(\mathbf{k})$. This suggests that the
%2-band Hubbard model\cite{yao,raghu} which includes only $d_{yz}$ orbital and $d_{zx}$
%orbital of a Fe atom should be irrelevant to describe the electronic states of
%iron-based superconductors from the point of view of the weak coupling theory\cite{kuroki,nomura_1}.  We see that the gap function for $s$-wave symmetry
%changes its sign between the hole pockets and the electron pockets and
%there are not any nodes on the Fermi surfaces.
%\cite{mazin,kuroki,nomura_1}. On the other hand, the gap function for
%$d_{xy}$-wave symmetry has line nodes on both the hole pockets and electron
%pockets. 
%%%%%%%%%%%%%%%%%%%%%%%%%%%%%% PHASE DIAGRAM %%%%%%%%%%%%%%%%%%%%%%%%%%%%%%%%%%%%%%%%%
\begin{figure}[t]
\begin{center} 
\includegraphics[width=6.0cm]{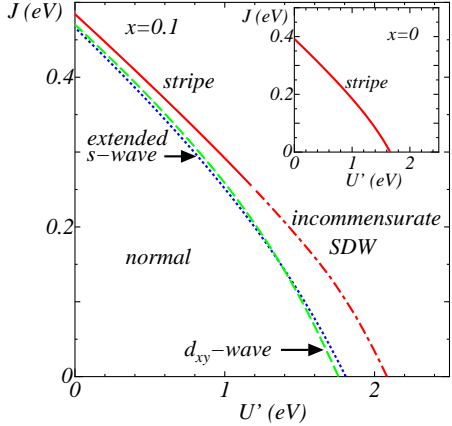}
\end{center}
\vspace{-5mm}
\caption{(Color online) The phase diagram on $U'$-$J$ plane for $x=0.1$ and $T=0.02\mathrm{eV}$. 
The lines represent the
 instabilities for the stripe antiferromagnetic order (solid), the
 incommensurate SDW (dot-dashed), the extended $s$-wave
 superconductivity (dotted) and the
 $d_{xy}$-wave superconductivity (dashed), respectively. The inset shows the magnetic phase diagram
 for $x=0$. \label{phasediagram}}
\end{figure}
%%%%%%%%%%%%%%%%%%%%%%%%%%%%%%%%%%%%%%%%%%%%%%%%%%%%%%%%%%%%%%%%%%%%%%%%%%%%%%%%%%%%%

The phase diagram on $U'$-$J$ plane for $x=0.1$ and $T=0.02\mathrm{eV}$ is shown
in Fig. \ref{phasediagram}, where the magnetic instability is determined by
the divergence of the spin susceptibility and the superconducting
instability is determined by $\lambda=1$ as mentioned before. For $J\simg0.25 \mathrm{eV}$,
the stripe-type antiferromagnetic order with $\mathbf{Q}=(\pi,\pi)$
appears, while, for $J\siml0.25\mathrm{eV}$, the incommensurate SDW
(ISDW) with $\mathbf{Q}\sim(\pi,\pi)$ appears (see also Figs. \ref{veff}
(a) and (b)). It is noted that we only observe the stripe-type
antiferromagnetic order for $x=0$ as shown in the inset in
Fig. \ref{phasediagram}. The extended
$s$-wave pairing is realized near the stripe-type antiferromagnetic
order for $J\simg0.15\mathrm{eV}$, where the spin fluctuation with
$\mathbf{Q}=(\pi,\pi)$ is enhanced as shown in Fig. \ref{veff} (b). On
the other hand, the $d_{xy}$-wave pairing is
realized near the ISDW for $J\siml0.15\mathrm{eV}$, where the spin
fluctuation with $\mathbf{Q}\sim(\pi,\pi)$ is enhanced as shown in
Fig. \ref{veff} (a). 

In summary, we investigated the pairing symmetry of the two-dimensional
16-band $d$-$p$ model by using the
RPA and obtained the phase diagram on the $U'$-$J$ plane for $x=0.1,
T=0.02\mathrm{eV}$. For  $J\simg0.15\mathrm{eV}$, the most favorable pairing is
extended $s$-wave symmetry whose order parameter changes
its sign between the hole pockets and the electron pockets, while for $J\siml0.15\mathrm{eV}$, it is $d_{xy}$-wave
symmetry. Then, the effect of the Hund's coupling $J$ is crucial to
realize the extended $s$-wave pairing in the pure system without
impurities. According to
the recent experiment of very weak $T_c$-suppression by Co-impurities\cite{kawabata_1}, we suppose that the $d_{xy}$-wave pairing is
suppressed by pair breaking effect
and the extended $s$-wave paring is realized in real materials\cite{senga}. 
\section*{Acknowledgment}
The authors thank M. Sato, H. Kontani and K. Kuroki for useful
comments and discussions. This work was partially supported by the
Grant-in-Aid for Scientific Research from the Ministry of Education,
Culture, Sports, Science and Technology. 

\end{document}